\documentclass[11pt]{article}

\usepackage[top=50mm, bottom=50mm, left=50mm, right=50mm]{geometry}
\usepackage{lineno}
\usepackage{natbib}
\usepackage{amssymb}
\usepackage{amsmath}
\usepackage{amsthm}
\usepackage{epsfig}
\usepackage{graphicx}
\usepackage{graphics}
\usepackage{float}
\usepackage{subfigure}
\usepackage{multirow}
\usepackage{color}
\usepackage{lineno}
\usepackage{fullpage}
\usepackage[normalem]{ulem}
\usepackage{makeidx}
\usepackage{xspace}
\usepackage{wrapfig}
\usepackage{algorithm}
\usepackage{verbdef}

\makeindex

\newtheorem{prop}{Proposition}
\newtheorem{Lemma}{Lemma}

\numberwithin{equation}{section}

\renewcommand{\baselinestretch}{2.0}

\begin{document}

\title{A cautionary note on the Hanurav-Vijayan sampling algorithm}

\author{
{Guillaume Chauvet\footnote{Univ Rennes, ENSAI, CNRS, IRMAR - UMR 6625, F-35000 Rennes, France, email: chauvet@ensai.fr}}
}

\maketitle
%\linenumbers
\setcounter{page}{0}

\maketitle

\begin{abstract}
\noindent We consider the Hanurav-Vijayan sampling design, which is the default method programmed in the \verb"SURVEYSELECT" procedure of the \verb"SAS" software. We prove that it is equivalent to the Sunter procedure, but is capable of handling any set of inclusion probabilities. We prove that the Horvitz-Thompson estimator is not generally consistent under this sampling design. We propose a conditional Horvitz-Thompson estimator, and prove its consistency under a non-standard assumption on the first-order inclusion probabilities. Since this assumption seems difficult to control in practice, we recommend not to use the Hanurav-Vijayan sampling design.
\end{abstract}

\section{Introduction} \label{sec0}

\noindent The Hanurav-Vijayan method \citep{vij:68} makes it possible to select a sample with probabilities proportional to size. This is the default method programmed in the \verb"SURVEYSELECT" procedure of the \verb"SAS" software for unequal probability sampling. It is therefore routinely used, see for example \citet{lan:fau:les:03,kul:kar:saa:kuu:07,myr:07,jan:pro:she:10,zha:11,cha:val:20,xio:hig:20}. \\

\noindent This sampling algorithm has a number of interesting features. The procedure is of fixed-size, the required first-order inclusion probabilities are exactly respected, and the second-order inclusion probabilities are strictly positive and may be computed. However, the statistical properties of estimators arising from the Hanurav-Vijayan method remain poorly studied, which may be due to the fairly intricate description of the method. This is the purpose of this paper. \\

\noindent After describing the main notations and assumptions in section \ref{sec1}, we present the Hanurav-Vijayan method in section \ref{sec2}. We prove that it is equivalent to the so-called Sunter sequential procedure, but that it can handle any set of first-order inclusion probabilities. The consistency of the Horvitz-Thompson estimator is studied in Section \ref{sec4}. In particular, we prove that the Horvitz-Thompson is not generally consistent under the Hanurav-Vijayan method, unless a non-standard condition on the first-order inclusion probabilites is respected. This condition requires that the $n$ largest inclusion probabilities are very close to each other. A conditional Horvitz-Thompson estimator is suggested in Section \ref{sec5}, and its consistency is established under a weaker condition. The results of the simulation study in section \ref{sec6} support our findings. We conclude in section \ref{sec7}.

\section{Notation} \label{sec1}

\noindent We consider a finite population $U=\{1,\ldots,N\}$. Denote by $\pi=(\pi_1,\ldots,\pi_N)^{\top}$ a vector of probabilities, with $0 < \pi_k < 1$ for any unit $k \in U$, and with $n = \sum_{k \in U}\pi_k$ the expected sample size. We suppose that the population $U$ is ordered with respect to the inclusion probabilities, i.e.
    \begin{eqnarray} \label{sec1:eq1}
    \pi_1 \leq \ldots \leq \pi_N.
    \end{eqnarray}
We note $\pi_l^+=\sum_{k=1}^l \pi_k$ for the cumulated inclusion probabilities up to unit $l$. A random sample is selected in $U$ by means of a without-replacement sampling design with parameter $\pi$, i.e. such that $E(I)=\pi$, where
    \begin{eqnarray} \label{sec1:eq2}
    I & = & (I_1,\ldots,I_N)^{\top}
    \end{eqnarray}
is the vector of sample membership indicators. We are interested in the estimation of the total $t_y=\sum_{k \in U} y_k$ for a variable of interest $y_k$.\\

\noindent Throughout the paper, we will consider the following assumptions:
\begin{itemize}
  \item[VA1:] There exists some constant $C_1$ such that:
    \begin{eqnarray*} \label{eq1:VA1}
      \frac{1}{N} \sum_{k \in U} y_k^2 & \leq & C_1.
    \end{eqnarray*}
  \item[SD1:] We have $n \to \infty$ as $N \to \infty$, and there exists some constant $f \in ]0,1[$ such that $N^{-1}n \to f$. There exists some constants $0<\lambda_1 \leq \Lambda_1$ such that for any $k \in U$:
    \begin{eqnarray*} \label{eq1:SD1}
      \lambda_1 \frac{n}{N} \leq & \pi_k & \leq \Lambda_1 \frac{n}{N}.
    \end{eqnarray*}
  \item[SD2:] There exists some function $h(n,N) \to 0$ such that
    \begin{eqnarray}
    \max_{i=1,\ldots,n-1} \{\pi_{N-n+i+1}-\pi_{N-n+i}\} & \leq & h(n,N).
    \end{eqnarray}
\end{itemize}

\noindent The assumption (VA1) is related to the variable of interest, which is assumed to have a finite moment of order 2. The assumption (SD1) is related to the sampling design, and also defines the asymptotic framework. It is assumed that all the first-order inclusion probabilities are of order $n/N$. The assumptions (VA1) and (SD1) are standard. The assumption (SD2) is more unusual, and states that the $n$ largest inclusion probabilities are sufficiently close to each other. Note that from the identity
    \begin{eqnarray*}
    \frac{1}{n-1} \sum_{i=1}^{n-1} \{\pi_{N-n+i+1}-\pi_{N-n+i}\} & = & \frac{\pi_N-\pi_{N-n+1}}{n-1},
    \end{eqnarray*}
the mean value of these differences is of order $N^{-1}$ under assumption (SD1). It would therefore seem natural to use $h(n,N)=N^{-1}$ in assumption (SD2). In any case, we prove in Section \ref{sec4} that under the Hanurav-Vijayan sampling process, $h(n,N)$ needs to be of smaller order for the Horvitz-Thompson estimator to be generally consistent.

\section{Hanurav-Vijayan procedure} \label{sec2}

\noindent The sampling algorithm proposed by \cite{vij:68} is a generalization of a procedure by \cite{han:67}. \cite{vij:68} considered the specific case of unequal probability sampling with probabilities proportional to size. The description in Algorithm \ref{algo:1} is more general, since it can be applied for any set $\pi$ of inclusion probabilities. We have also simplified the presentation, to express the intermediary quantities needed in the sampling process in terms of the inclusion probabilities only. \\

\begin{algorithm}
\caption{Hanurav-Vijayan procedure: draw by draw algorithm} \label{algo:1}
Phase 1:
\begin{itemize}
    \item Select an integer $n'$ with probabilities
        \begin{eqnarray} \label{sec1:eq3}
        \delta_{i} & = & (\pi_{N-n+i+1}-\pi_{N-n+i}) \frac{\pi_{N-n}^+ + i \pi_{N-n+1}}{\pi_{N-n}^+} \textrm{ for } i \in \{1,\ldots,n\},
        \end{eqnarray}
    where $\pi_{N+1}=1$. We note $N'=N-(n-n')$.
    \item Take $\pi(0)=\{\pi_1(0),\ldots,\pi_N(0)\}^{\top}$, such that
        \begin{eqnarray} \label{sec1:eq4}
        \pi_k(0) & = & \left\{
        \begin{array}{cc}
        \frac{n' \pi_k}{\pi_{N-n}^+ + n' \pi_{N-n+1}} & \textrm{ if } k \leq N-(n-1), \\
        \frac{n' \pi_{N-n+1}}{\pi_{N-n}^+ + n' \pi_{N-n+1}} & \textrm{ if } N-(n-1)<k\leq N', \\
        1 & \textrm{ if } k>N'.
        \end{array}
        \right.
        \end{eqnarray}
\end{itemize}

Phase 2: In the population $U'=\{1,\ldots,N'\}$, select a sample of size $n'$ as follows:
    \begin{itemize}
        \item Initialize with $i_0=0$.
        \item For $j=1,\ldots,n'$, select one unit $i_j$ from $\{i_{j-1}+1,\ldots,N-n+j\}$ with probabilities proportional to
            \begin{eqnarray} \label{sec1:eq5}
            a_{i_{j-1}+1}^j & = & \frac{n'-j+1}{n'} \pi_{i_{j-1}+1}(0)
            \end{eqnarray}
        for unit $i_{j-1}+1$ and
            \begin{eqnarray} \label{sec1:eq6}
            a_k^j & = & \prod_{l=i_{j-1}+1}^{k-1} \left\{1-(n'-j) \frac{\pi_l(0)}{n'-\pi_{l}^{+}(0)} \right\} \times \frac{n'-j+1}{n'} \pi_{k}(0)
            \end{eqnarray}
        for $k=i_{j-1}+2,\ldots,N-n+j$, where $\pi_{l}^{+}(0)=\sum_{k=1}^l \pi_k(0)$.
    \end{itemize}
The final sample is: $S=\{i_1,\ldots,i_{n'},N'+1,\ldots,N\}$.
\end{algorithm}

\noindent The Hanurav-Vijayan procedure is split into two phases. During the first phase, an integer $n'$ is randomly selected in $\{1,\ldots,n\}$ and a new vector $\pi(0)$ of inclusion probabilities is obtained. The $n-n'$ units with the larger inclusion probabilities ($k>N-n+n'$) are selected, while the $n'$ remaining units with the larger inclusion probabilities ($N-n+1<k\leq N-n+n'$) are given the same value $\frac{n' \pi_{N-n+1}}{\pi_{N-n}^+ + n' \pi_{N-n+1}}$. During the second phase, a sample of size $n'$ is selected among the remaining units through a draw by draw procedure. The algorithm is of fixed size by construction. We have $E\{\pi(0)\}=\pi$, and conditionally on $\pi(0)$ the sampling in $U'$ is performed with inclusion probabilities $\pi(0)$ \citep[see][Theorem 1]{vij:68}. Therefore, the original set of inclusion probabilities $\pi$ is exactly respected. We denote by $\pi_{kl}(0) = E\{I_k I_l|\pi(0)\}$ the second-order inclusion probability of units $k,l \in U'$ during Phase 2, conditionally on $\pi(0)$. \\

\noindent The random rounding in Phase 1 ensures that $\pi_{N-n+1}(0)=\pi_{N-n+2}(0)=\ldots=\pi_{N-n+n'}(0)$, which is necessary for the suitability of the draw by draw procedure in Phase 2. This is an early example of the splitting method later theorized by \cite{dev:til:98}. Note that if $\pi_{N-n+1}=\pi_N$, we obtain $n'=n$ with probability $1$ and $\pi(0)=\pi$, which means that Phase 1 is not needed. For example, this occurs when sampling with equal probabilities, in which case the Hanurav-Vijayan procedure is equivalent to simple random sampling. \\

\noindent The second phase of the Hanurav-Vijayan algorithm may be more simply implemented in terms of a sequential procedure, presented in Algorithm \ref{algo:2}. Proposition \ref{prop1} states that both sampling algorithms are equivalent. The proof is given in Appendix \ref{appA}. The second phase of Algorithm \ref{algo:2} is a generalization of the selection-rejection method \citep{fan:mul:rez:62} for unequal probability sampling, known as the Sunter procedure \citep{sun:77,sun:86}. The Sunter procedure is known to be non-exact, in the sense that it cannot be directly applied to any set of inclusion probabilities \citep[e.g.][Section 6.2.8]{til:06}. The first phase of the Hanurav-Vijayan algorithm makes the Sunter algorithm applicable in full generality. It is remarkable that this solution was proposed ten years before the sequential procedure was  introduced by \cite{sun:77}. Another possible generalization is proposed in \cite{dev:til:98}. \\

\begin{algorithm}
\caption{Hanurav-Vijayan-Sunter procedure: sequential algorithm} \label{algo:2}
Phase 1:
\begin{itemize}
    \item Select an integer $n'$ with probabilities
        \begin{eqnarray*}
        \delta_{i} & = & (\pi_{N-n+i+1}-\pi_{N-n+i}) \frac{\pi_{N-n}^+ + i \pi_{N-n+1}}{\pi_{N-n}^+} \textrm{ for } i \in \{1,\ldots,n\},
        \end{eqnarray*}
    where $\pi_{N+1}=1$. We note $N'=N-(n-n')$.
    \item Take $\pi(0)=\{\pi_1(0),\ldots,\pi_N(0)\}^{\top}$, such that
        \begin{eqnarray*}
        \pi_k(0) & = & \left\{
        \begin{array}{cc}
        \frac{n' \pi_k}{\pi_{N-n}^+ + n' \pi_{N-n+1}} & \textrm{ if } k \leq N-(n-1), \\
        \frac{n' \pi_{N-n+1}}{\pi_{N-n}^+ + n' \pi_{N-n+1}} & \textrm{ if } N-(n-1)<k\leq N',  \\
        1 & \textrm{ if } k>N'.
        \end{array}
        \right.
        \end{eqnarray*}
\end{itemize}

Phase 2: In the population $U'=\{1,\ldots,N'\}$, select a sample of size $n'$ as follows. Initialize with $n_0=0$. % and $\pi^{+}(0)=0$.
 For $t=1,\ldots,N'-1$:
    \begin{itemize}
        \item take $I_t=1$ with probability $\pi_t(t-1)$, and $n_t=n_{t-1}+I_t$,
        \item compute $\pi(t)=\{\pi_1(t),\ldots,\pi_{N'}(t)\}^{\top}$ such that
                \begin{eqnarray} \label{sec1:eq7}
                \pi_k(t) & = & \left\{
                \begin{array}{ll}
                    \pi_k(t-1) & \textrm{ if } k \leq t-1,  \\
                     I_t & \textrm{ if } k =t, \\
                     (n'-n_t) \frac{\pi_k(0)}{n'-\pi_{t}^+(0)} & \textrm{ if } k >t,
                \end{array}
                \right.
                \end{eqnarray}
            where $\pi_{t}^+(0)=\sum_{k=1}^{t} \pi_{k}(0)$.
    \end{itemize}
The vector of sample membership indicators is $I=\{\pi_1(N'-1),\ldots,\pi_{N'}(N'-1),1,\ldots,1\}^{\top}$.
\end{algorithm}

\begin{prop} \label{prop1}
Algorithms \ref{algo:1} and \ref{algo:2} lead to the same sampling design.
\end{prop}

\section{Horvitz-Thompson estimator} \label{sec4}

\noindent In this section, we are interested in the Horvitz-Thompson (HT) estimator
    \begin{eqnarray} \label{sec4:eq1}
    \hat{t}_{y\pi} & = & \sum_{k \in S} \frac{y_k}{\pi_k}.
    \end{eqnarray}
We make use of the indicator
    \begin{eqnarray} \label{sec4:eq2}
    D_1(\pi) & = & \frac{1}{n} \sum_{i=1}^{n-1} (n-i) (\pi_{N-n+i+1} - \pi_{N-n+i}),
    \end{eqnarray}
which can be seen as a measure of distance between the $n$ largest inclusion probabilities. We also use the notation
    \begin{eqnarray} \label{sec4:eq3}
    \xi(0) & = & \sum_{k \in U} \frac{y_k}{\pi_k} \left\{\pi_k(0)-\pi_k \right\}.
    \end{eqnarray}
We have
    \begin{eqnarray} \label{sec4:eq4}
    V(\hat{t}_{y\pi}) & = & VE\left\{\hat{t}_{y\pi}|\pi(0)\right\}+EV\left\{\hat{t}_{y\pi}|\pi(0)\right\} \\
                      & \geq & VE\left\{\hat{t}_{y\pi}|\pi(0)\right\} = V\left\{\xi(0) \right\} = \sum_{i=1}^n \delta_i E\left\{\xi(0)^2|n'=i\right\} \nonumber \\
                      & \geq & \delta_n \left[\left\{\frac{n}{\pi_{N-n}^+ +n\pi_{N-n+1}}-1\right\} \sum_{k=1}^{N-n} y_k  +
                      \sum_{k=N-n+1}^{N} y_k \left\{\frac{n\pi_{N-n+1}}{\pi_k(\pi_{N-n}^+ +n\pi_{N-n+1})}-1 \right\} \right]^2, \nonumber
    \end{eqnarray}
where the last line in (\ref{sec4:eq4}) is obtained by keeping the case $i=n$ only. The inequality (\ref{sec4:eq4}) gives the basic idea of why the HT-estimator may be inconsistent. The term $V\left\{\xi(0) \right\}$ is due to the randomization in Phase 1, which is needed for the suitability of the sampling in Phase 2. In some cases, this variability does not vanish as $n \to \infty$, as stated in Proposition \ref{prop:incons}. \\

\begin{prop} \label{prop:incons}
Suppose that assumption (SD1) holds, and that there exists some constant $0<\lambda_2$ such that
    \begin{eqnarray} \label{sec4:eq5}
    \lambda_2 \frac{n}{N} & \leq & D_1(\pi).
    \end{eqnarray}
Suppose that there exists some constants $c_2>0$ and $C_2$ such that
    \begin{eqnarray} \label{prop:incons:eq1}
    c_2 \leq \frac{1}{N-n}  \left|\sum_{k=1}^{N-n} y_k \right|
    & \textrm{ and } &
    \frac{1}{n} \sum_{k=N-n+1}^{N} |y_k| \leq C_2.
    \end{eqnarray}
If
    \begin{eqnarray} \label{prop:incons:eq2}
    \frac{c_2}{C_2} & > & \frac{1}{\lambda_2(1-f)} \frac{\Lambda_1}{\lambda_1} \left(1+\frac{1}{\lambda_1} \right),
    \end{eqnarray}
where the constants $f$, $\lambda_1$ and $\Lambda_1$ are defined in assumption (SD1), then there exists some constant $C>0$ such that:
    \begin{eqnarray} \label{prop:incons:eq3}
    V(N^{-1} \hat{t}_{y\pi}) & \geq & (1-\pi_N) C N^{-2} n^2.
    \end{eqnarray}
\end{prop}

\noindent Proposition \ref{prop:incons} states that if the indicator $D_1(\pi)$ is too large, we can always find variables of interest satisfying assumption (VA1) and such that the HT-estimator is not consistent, since the second term in the right-hand side of (\ref{prop:incons:eq3}) is bounded away from $0$. The proof is given in Appendix \ref{app:incons}. The ratio $c_2/C_2$ may be thought of as a measure of balance of the total $t_y$ between the $N-n$ first units and the $n$ last units: the HT-estimator is not consistent if the total $t_y$ is too highly concentrated on the $N-n$ first units. \\

\begin{prop} \label{prop:cons}
Suppose that assumptions (VA1) and (SD1) hold, and that assumption (SD2) holds with $h(n,N)=o(N^{-1})$. Then
    \begin{eqnarray} \label{prop:cons:eq1}
    V(N^{-1} \hat{t}_{y\pi}) & = & o(1).
    \end{eqnarray}
\end{prop}

\noindent Proposition \ref{prop:cons} states that the HT-estimator is consistent if the $n$ largest inclusion probabilities are sufficiently close, namely if assumption (SD2) holds with $h(n,N)=o(N^{-1})$. This assumption can not be dropped. For example, if there is a constant lag of order $N^{-1}$ between these probabilities, namely if there exists some constant $0<\lambda$ such that
    \begin{eqnarray*}
    \pi_{N-n+i+1}-\pi_{N-n+i} & = & \frac{\lambda}{N},
    \end{eqnarray*}
then $D_1(\pi)=\frac{\lambda}{2} \frac{n-1}{N}$. Therefore, equation (\ref{sec4:eq5}) in Proposition  \ref{prop:incons} holds, and there are some variables of interest such that (VA1) holds but the HT-estimator is not consistent. \\

\noindent From a look at the proof of Proposition \ref{prop:cons}, the assumption (SD2) is needed to control the term $VE(\hat{t}_{y\pi}|\pi(0))$, which is due to the first phase in Algorithm \ref{algo:1}. To remove this variability, it is possible to work conditionally on $\pi(0)$. This is the purpose of the next section.

\section{Conditional Horvitz-Thompson estimator} \label{sec5}

\noindent We are interested in the conditional Horvitz-Thompson (CHT) estimator, defined as
    \begin{eqnarray} \label{sec5:eq1}
    \hat{t}_{y\pi}(0) & = & \sum_{k \in S} \frac{y_k}{\pi_k(0)}.
    \end{eqnarray}
This estimator makes use of the set of inclusion probabilities $\pi(0)$ obtained after Phase 1 of Algorithm \ref{algo:1}. It may be rewritten as
    \begin{eqnarray} \label{sec5:eq2}
    \hat{t}_{y\pi}(0) & = & \sum_{k \in U'} \frac{y_k}{\pi_k(0)} I_k + \sum_{k>N'} y_k,
    \end{eqnarray}
which leads to
    \begin{eqnarray} \label{sec5:eq3}
    E\left\{\hat{t}_{y\pi}(0)|\pi(0)\right\} & = & \sum_{k \in U'} y_k+\sum_{k>N'} y_k = t_y.
    \end{eqnarray}
This is therefore an unbiased estimator for $t_y$, conditionally on $\pi(0)$. \\

\begin{prop} \label{prop3}
Suppose that assumptions (VA1) and (SD1) hold, and that assumption (SD2) holds with $\displaystyle h(n,N)=o\left(\frac{1}{\ln(n)}\right)$. Then
    \begin{eqnarray} \label{prop:cons:eq2}
    V\{N^{-1} \hat{t}_{y\pi}(0)\} & = & o(1).
    \end{eqnarray}
If the assumption (SD2) holds with $\displaystyle h(n,N)=O\left(\frac{1}{n\ln(n)}\right)$, then the CHT-estimator is $\sqrt{n}$-consistent.
\end{prop}

\noindent The proof of Proposition \ref{prop3} is given in Appendix \ref{appB}. We clearly need a weaker assumption on the difference of the largest inclusion probabilities. Anyway, we need these differences to be no greater than $O\left(\frac{1}{n\ln(n)}\right)$ to ensure the usual $\sqrt{n}$-consistency, which is still demanding. \\

\noindent Another advantage of the CHT-estimator is that the variance may be easily estimated. From the corollary of Theorem 1 in \citet{vij:68}, there is an explicit expression for the conditional second-order inclusion probabilities, which is restated in Proposition \ref{prop4b}. Note that an incorrect factor of $\frac{1}{2}$ was indicated in equation (\ref{prop4b:eq1}) by \citet{vij:68}, see \citet{cha:vos:88}.

\begin{prop} (Vijayan, 1968)  \label{prop4b}
For $k=1,\ldots,N'$, we note
    \begin{eqnarray*}
    p_k(0) = \frac{\pi_k(0)}{n'} & \textrm{ and } & P_k(0) = \frac{\pi_k(0)}{n'-\pi_k^+(0)}.
    \end{eqnarray*}
For $k<l=1,\ldots,N'$, we have
    \begin{eqnarray} \label{prop4b:eq1}
    \pi_{kl}(0) & = & n'(n'-1) \{1-P_1(0)\} \ldots \{1-P_{k-1}(0)\} P_k(0) p_l(0).
    \end{eqnarray}
For $k=1,\ldots,N'$ and $l=N'+1,\ldots,N$, we have
    \begin{eqnarray} \label{prop4b:eq2}
    \pi_{kl}(0) & = & \pi_k(0).
    \end{eqnarray}
For $k<l=N'+1,\ldots,N$, we have
    \begin{eqnarray} \label{prop4b:eq3}
    \pi_{kl}(0) & = & 1.
    \end{eqnarray}
\end{prop}

\noindent The second-order inclusion probabilities $\pi_{kl}(0)$ are strictly positive, and satisfy the Sen-Yates-Grundy conditions \citep[see][Theorem 3]{vij:68}. Therefore, the Sen-Yates-Grundy variance estimator is unbiased and takes positive values only. In Theorem 2 of \cite{vij:68}, these probabilities are averaged to obtain the unconditional second-order inclusion probabilities for the HT estimator. However, this involves computing the $\pi_{kl}(0)$'s for each of the $n$ possible cases for the integer $n'$, which is cumbersome if $n$ is large.

\section{Simulation study} \label{sec6}

\noindent We conduct a simulation study to illustrate the properties of the Horvitz-Thompson (HT) estimator and of the conditional Horvitz-Thompson (CHT) estimator. The set-up is inspired from \citet{cha:20}. We generate $2$ populations of size $N$, each consisting of an auxiliary variable $x$ and $4$ variables of interest $y_1,\ldots,y_4$. The $x$-values are generated according to the model
    \begin{eqnarray}
    x_k & = & \alpha + \eta_k.
    \end{eqnarray}
In the first population, we use $\alpha=8$ and $\eta_k$ is generated according to a Gamma distribution with shape and scale parameters $4$ and $0.5$. In the second population, we use $\alpha=7$ and $\eta_k$ is generated according to a log-normal distribution with parameters $1.0$ and $0.35$. This leads to a mean of approximately $10$ and a standard deviation of approximately $1$ for the variable $x$ in both populations. \\

\noindent  Given the $x$-values, the variables of interest are generated according to the following models:
    \begin{eqnarray} \label{sec6:eq1}
      \verb"linear": y_{1k} & = & \alpha_{10} + \alpha_{11} (x_k - \mu_x) + \sigma_1~\epsilon_k, \nonumber \\
      \verb"quadratic": y_{2k} & = & \alpha_{20} + \alpha_{21} (x_k - \mu_x)^2 + \sigma_2~\epsilon_k, \\
      \verb"exponential": y_{3k} & = & \exp\{\alpha_{30} + \alpha_{31}(x_k - \mu_x)\}+\sigma_3~\epsilon_k, \nonumber \\
      \verb"bump": y_{4k} & = & \alpha_{40} + \alpha_{41} (x_k - \mu_x)^2 - \alpha_{42} \exp\left\lbrace-\alpha_{43}( x_k - \mu_x)^2\right\rbrace + \sigma_4~\epsilon_k, \nonumber
    \end{eqnarray}
where $\mu_x$ is the population mean of $x$, and $\epsilon_k$ follows a standard normal distribution. The parameters are chosen in order to obtain a mean of approximately $20$ and a standard deviation of approximately $3$ for each variable of interest. \\

\noindent In each population, we compute inclusion probabilities proportional to $x$, according to the formula
    \begin{eqnarray} \label{sec6:eq2}
      \pi_k & = & n \frac{x_k}{\sum_{l \in U} x_l}.
    \end{eqnarray}
We use ten different population sizes, ranging from $N=2,000$ to $N=20,000$, and a sampling fraction of $20 \% $ for each population. This leads to sample sizes ranging from $n=400$ to $n=4,000$. For example, when $N=20,000$, the inclusion probabilities range between $0.16$ and $0.32$ when $x$ is generated by means of the Gamma distribution, and between $0.15$ and $0.38$ when $x$ is generated by means of the log-normal distribution. \\

\begin{figure}[htbp]
    \centering \includegraphics[height=6in,width=6in]{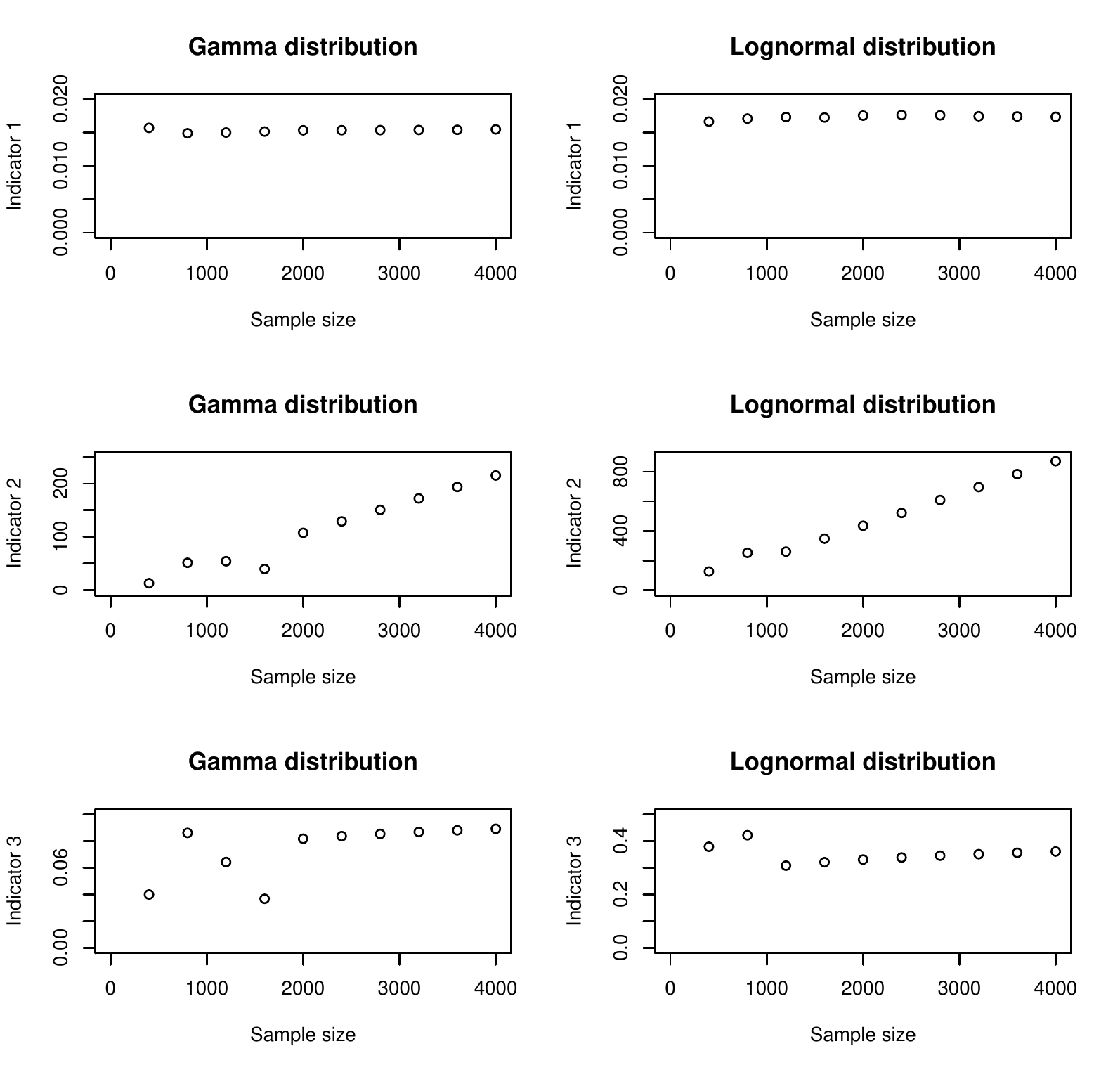}
    \caption{Indicators $D_1(\pi)$, $D_2(\pi)$ and $D_3(\pi)$ in function of the sample size $n$ with an auxiliary variable generated according to a gamma distribution (lhs) and by a lognormal distribution (rhs)}\label{fig1}
\end{figure}

\noindent We consider the indicator $D_1(\pi)$ defined in equation (\ref{sec4:eq2}), and the additional indicators
    \begin{eqnarray*}
    D_2(\pi) & = & N \times \max_{i=1,\ldots,n-1} \{\pi_{N-n+i+1}-\pi_{N-n+i}\}, \\
    D_3(\pi) & = & \ln(n) \times \max_{i=1,\ldots,n-1} \{\pi_{N-n+i+1}-\pi_{N-n+i}\}.
    \end{eqnarray*}
If the assumption (SD2) is respected with $h(n,N)=o(N^{-1})$ (see Proposition \ref{prop:incons}), then $D_1(\pi)=o(1)$ and $D_2(\pi)=o(1)$, and they should therefore tend to $0$ as $n$ increases. If the assumption (SD2) is respected with $h(n,N)=o(1/\ln(n))$ (see Proposition \ref{prop:cons}), then $D_3(\pi)=o(1)$ and $D_3(\pi)$ should therefore tend to $0$ as $n$ increases. We have plotted these indicators in terms of the sample size $n$ in Figure \ref{fig1}. Neither of them decreases as $n$ increases. The indicator $D_1(\pi)$ is approximately constant, and so is the indicator $D_3(\pi)$ for large sample sizes ($n \geq 2,000$). The indicator $D_3(\pi)$ is clearly increasing with $n$. This supports the apparent difficulties for controlling the closeness of the largest inclusion probabilities via the assumption (SD2). \\

\noindent We consider the estimation of the population mean $\mu_y = N^{-1} \sum_{k \in U} y_k$. We select $B=10,000$ samples by means of the HVS sampling algorithm. For each sample and each variable of interest, we consider the population mean $\mu_y = N^{-1} \sum_{k \in U} y_k$. We compute the Horvitz-Thompson estimator of the mean $\hat{\mu}_{y\pi} = N^{-1} \hat{t}_{y\pi}$, and the conditional estimator of the mean $\hat{\mu}_{y\pi}(0) = N^{-1} \hat{t}_{y\pi}(0)$. For each estimator $\hat{\mu}_y$ and for a given sample size $n$, we compute the Monte-Carlo variance
    \begin{eqnarray}
    V_{MC,n}(\hat{\mu}_{y}) & = & \frac{1}{B} \sum_{b=1}^B \left\{\hat{\mu}_y(s_b)-\frac{1}{B} \sum_{c=1}^B \hat{\mu}_y(s_c)\right\}^2,
    \end{eqnarray}
with $\hat{\mu}_y(s_b)$ the estimator of the mean computed on the $b$-th sample. We also compute the Monte-Carlo variance ratio
    \begin{eqnarray}
    {RV}_{MC,n}(\hat{t}_{y}) & = & \frac{V_{MC,n}(\hat{t}_{y})}{V_{MC,n-400}(\hat{t}_{y})}.
    \end{eqnarray}
If the estimator $\hat{t}_{y}$ is consistent, the Monte-Carlo variance is expected to decrease as the sample size increases, and the Monte-Carlo variance ratios should be lower than $1$. \\

\noindent The simulation results for the HT-estimator are presented in Table \ref{tab1}. In 17 out of 72 cases the variance ratio ${RV}_{MC,n}$ is greater than $1$, indicating that the variance increases as $n$ increases. In Population 1, the behavior of $\hat{\mu}_{y\pi}$ is particularly poor for \verb"quadratic", since the variance is of the same order with $n=400$ ($27.84 \times 10^{-3}$) and $n=4,000$ ($19.41 \times 10^{-3}$). In Population 2, the behavior of $\hat{\mu}_{y\pi}$ is particularly poor for \verb"exponential", since the variance is of the same order with $n=400$ ($26.77 \times 10^{-3}$) and $n=4,000$ ($21.75 \times 10^{-3}$). This supports the results in Section \ref{sec4}. \\

\noindent The simulation results for the CHT-estimator are presented in Table \ref{tab2}. The variance ratio ${RV}_{MC,n}$ is lower than $1$ in 69 out of 72 cases, ${RV}_{MC,n}$ being lower than $1.03$ in the three remaining cases.
In almost all cases, the variance obtained with $n=4,000$ is roughly one tenth of the variance obtained with $n=400$, as could be expected. This supports the consistency result obtained in Proposition \ref{prop3}, even if the assumption (SD2) in Proposition \ref{prop:cons} is not exactly respected (see the indicator $D_3(\pi)$ plotted in Figure \ref{fig1}). We note that the CHT-estimator is not necessarily more efficient than the HT-estimator. For \verb"linear", the variance of the HT-estimator is systematically lower.

\verbdef{\linear}{linear}
\verbdef{\quadratic}{quadratic}
\verbdef{\exponential}{exponential}
\verbdef{\bump}{bump}

{\renewcommand{\baselinestretch}{1.4}
\begin{table}[htb!]
\caption{Monte-Carlo variance ($V_{MC,n}$) and Monte-Carlo variance ratio (${RV}_{MC,n}$) for the Horvitz-Thompson (HT) estimator, for two populations and four variables of interest} \label{tab1}
	\begin{center}
	\begin{tabular}{|l|cccccccccc|} \hline
	    & \multicolumn{10}{|c|}{Population 1 (Gamma distribution)} \\ \hline
	Sample size & $400$ & $800$ & $1,200$ & $1,600$ & $2,000$ & $2,400$ & $2,800$ & $3,200$ & $3,600$ & $4,000$ \\ \hline
	    & \multicolumn{10}{|c|}{\linear} \\
$V_{MC,n}$ ($\times 10^{-3}$) & 8.10  & 4.37   & 2.65   & 1.99   & 1.64   & 1.30   & 1.18   & 0.92   & 0.83   & 0.80   \\
$({RV}_{MC,n})$  &       & (0.54) & (0.61) & (0.75) & (0.83) & (0.79) & (0.91) & (0.77) & (0.91) & (0.96) \\
	    & \multicolumn{10}{|c|}{\quadratic} \\
$V_{MC,n}$ ($\times 10^{-3}$) & 27.84 & 28.14  & 18.42  & 20.22  & 15.90  & 16.84  & 24.68  & 14.43  & 12.31  & 19.41  \\
$({RV}_{MC,n})$  &       & (1.01) & (0.65) & (1.10) & (0.79) & (1.06) & (1.47) & (0.58) & (0.85) & (1.58) \\
	    & \multicolumn{10}{|c|}{\exponential} \\
$V_{MC,n}$ ($\times 10^{-3}$) & 9.42  & 5.55   & 4.77   & 3.58   & 3.68   & 3.33   & 3.84   & 3.12   & 2.72   & 3.38   \\
$({RV}_{MC,n})$  &       & (0.59) & (0.86) & (0.75) & (1.03) & (0.90) & (1.15) & (0.81) & (0.87) & (1.24) \\
	    & \multicolumn{10}{|c|}{\bump} \\
$V_{MC,n}$ ($\times 10^{-3}$) & 37.23 & 16.40  & 12.31  & 9.45   & 7.11   & 6.47   & 5.51   & 4.33   & 4.23   & 3.92   \\
$({RV}_{MC,n})$  &       & (0.44) & (0.75) & (0.77) & (0.75) & (0.91) & (0.85) & (0.79) & (0.98) & (0.93) \\ \hline \hline
	    & \multicolumn{10}{|c|}{Population 2 (Log-normal distribution)} \\ \hline
	Sample size & $400$ & $800$ & $1,200$ & $1,600$ & $2,000$ & $2,400$ & $2,800$ & $3,200$ & $3,600$ & $4,000$ \\ \hline
	    & \multicolumn{10}{|c|}{\linear} \\
$V_{MC,n}$ ($\times 10^{-3}$) & 8.42  & 4.04   & 2.94   & 2.16   & 1.72   & 1.32   & 1.24   & 1.05   & 0.95   & 0.77   \\
$({RV}_{MC,n})$  &       & (0.48) & (0.73) & (0.74) & (0.80) & (0.77) & (0.94) & (0.85) & (0.90) & (0.81) \\
	    & \multicolumn{10}{|c|}{\quadratic} \\
$V_{MC,n}$ ($\times 10^{-3}$) & 37.61 & 36.57  & 24.70  & 37.25  & 27.30  & 17.26  & 19.51  & 29.01  & 24.15  & 27.77  \\
$({RV}_{MC,n})$  &       & (0.97) & (0.68) & (1.51) & (0.73) & (0.63) & (1.13) & (1.49) & (0.83) & (1.15) \\
	    & \multicolumn{10}{|c|}{\exponential} \\
$V_{MC,n}$ ($\times 10^{-3}$) & 26.77 & 27.11  & 20.49  & 28.29  & 22.09  & 15.00  & 16.08  & 23.31  & 19.20  & 21.75  \\
$({RV}_{MC,n})$  &       & (1.01) & (0.76) & (1.38) & (0.78) & (0.68) & (1.07) & (1.45) & (0.82) & (1.13) \\
	    & \multicolumn{10}{|c|}{\bump} \\
$V_{MC,n}$ ($\times 10^{-3}$) & 37.53 & 19.38  & 13.05  & 9.90   & 8.25   & 6.48   & 6.19   & 5.56   & 4.98   & 4.28   \\
$({RV}_{MC,n})$  &       & (0.52) & (0.67) & (0.76) & (0.83) & (0.78) & (0.95) & (0.90) & (0.90) & (0.86) \\ \hline
	\end{tabular}
	\end{center}
\end{table}
}

{\renewcommand{\baselinestretch}{1.4}
\begin{table}[htb!]
\caption{Monte-Carlo variance ($V_{MC,n}$) and Monte-Carlo variance ratio (${RV}_{MC,n}$) for the conditional Horvitz-Thompson (CHT) estimator, for two populations and four variables of interest} \label{tab2}
	\begin{center}
	\begin{tabular}{|l|cccccccccc|} \hline
	    & \multicolumn{10}{|c|}{Population 1 (Gamma distribution)} \\ \hline
	Sample size & $400$ & $800$ & $1,200$ & $1,600$ & $2,000$ & $2,400$ & $2,800$ & $3,200$ & $3,600$ & $4,000$ \\ \hline
	    & \multicolumn{10}{|c|}{\linear} \\
$V_{MC,n}$ ($\times 10^{-3}$) & 9.11  & 5.16   & 3.22    & 2.40   & 1.92   & 1.46   & 1.43   & 1.06   & 1.03   & 0.99   \\
$({RV}_{MC,n})$ &       & (0.57) & (0.62)  & (0.75) & (0.80) & (0.76) & (0.98) & (0.74) & (0.96) & (0.96) \\
	    & \multicolumn{10}{|c|}{\quadratic} \\
$V_{MC,n}$ ($\times 10^{-3}$) & 12.46 & 8.31   & 4.16    & 3.34   & 2.50   & 2.06   & 1.91   & 1.55   & 1.34   & 1.29   \\
$({RV}_{MC,n})$ &       & (0.67) & (0.50)  & (0.80) & (0.75) & (0.82) & (0.93) & (0.81) & (0.86) & (0.96) \\
	    & \multicolumn{10}{|c|}{\exponential} \\
$V_{MC,n}$ ($\times 10^{-3}$) & 10.21 & 5.80   & 3.68    & 2.73   & 2.22   & 1.67   & 1.65   & 1.23   & 1.19   & 1.13   \\
$({RV}_{MC,n})$ &       & (0.57) & (0.63)  & (0.74) & (0.81) & (0.75) & (0.99) & (0.75) & (0.97) & (0.95) \\
	    & \multicolumn{10}{|c|}{\bump} \\
$V_{MC,n}$ ($\times 10^{-3}$) & 37.07 & 30.31  & 12.62   & 10.69  & 7.22   & 6.53   & 5.40   & 4.32   & 4.34   & 3.96   \\
$({RV}_{MC,n})$ &       & (0.82) & (0.42)  & (0.85) & (0.67) & (0.91) & (0.83) & (0.80) & (1.01) & (0.91) \\ \hline \hline
	    & \multicolumn{10}{|c|}{Population 2 (Log-normal distribution)} \\ \hline
	Sample size & $400$ & $800$ & $1,200$ & $1,600$ & $2,000$ & $2,400$ & $2,800$ & $3,200$ & $3,600$ & $4,000$ \\ \hline
	    & \multicolumn{10}{|c|}{\linear} \\
$V_{MC,n}$ ($\times 10^{-3}$) & 10.80 & 5.11   & 3.50    & 2.50   & 2.01   & 1.69   & 1.47   & 1.29   & 1.29   & 0.91   \\
$({RV}_{MC,n})$ &       & (0.47) & (0.68)  & (0.71) & (0.81) & (0.84) & (0.87) & (0.88) & (1.00) & (0.70) \\
	    & \multicolumn{10}{|c|}{\quadratic} \\
$V_{MC,n}$ ($\times 10^{-3}$) & 16.10 & 7.50   & 4.95    & 3.57   & 2.98   & 2.32   & 2.07   & 2.08   & 1.83   & 1.27   \\
$({RV}_{MC,n})$ &       & (0.47) & (0.66)  & (0.72) & (0.83) & (0.78) & (0.89) & (1.01) & (0.88) & (0.69) \\
	    & \multicolumn{10}{|c|}{\exponential} \\
$V_{MC,n}$ ($\times 10^{-3}$) & 10.61 & 4.90   & 3.37    & 2.43   & 1.99   & 1.55   & 1.38   & 1.31   & 1.24   & 0.86   \\
$({RV}_{MC,n})$ &       & (0.46) & (0.69)  & (0.72) & (0.82) & (0.78) & (0.89) & (0.95) & (0.94) & (0.69) \\
	    & \multicolumn{10}{|c|}{\bump} \\
$V_{MC,n}$ ($\times 10^{-3}$) & 40.49 & 21.34  & 13.10   & 8.85   & 7.84   & 7.13   & 5.93   & 4.62   & 4.75   & 3.60   \\
$({RV}_{MC,n})$ &       & (0.53) & (0.61)  & (0.68) & (0.89) & (0.91) & (0.83) & (0.78) & (1.03) & (0.76) \\ \hline
	\end{tabular}
	\end{center}
\end{table}
}

\section{Conclusion} \label{sec7}

\noindent In this paper, we have studied the Hanurav-Vijayan sampling algorithm. We have proposed a sequential characterization of the method, making the link with Sunter's procedure. We have also shown that to ensure the consistency of the Horvitz-Thompson estimator, or of an alternative conditional Horvitz-Thompson estimator, we need to control the closeness between the largest inclusion probabilities. This seems rather difficult to achieve in practice. On the other hand, alternative unequal probability sampling methods programmed in the \verb"SURVEYSELECT" procedure lead to a consistent Horvitz-Thompson under the sole assumptions (VA1) and (SD1). This is the case for the Sampford method \citep{sam:67} or Chromy's method \citep{chr:79,cha:20}, for example. Therefore, we recommend that SAS users consider one of these two methods instead.

\section*{Acknowledgments} 

\noindent I would like to thank Todd Donahue for his careful reading of the manuscript. 

%\newpage

\bibliographystyle{apalike}
%\bibliography{biblio_gen}

\appendix

\section{Proof of Proposition \ref{prop1}} \label{appA}

\noindent Let $k_1,\ldots,k_{n'}$ denote the $n'$ units successively selected during Phase 2 of Algorithm \ref{algo:2}. It is sufficient to prove that their probability distribution is the same as that of the units $i_1,\ldots,i_{n'}$ successively selected during Phase 2 of Algorithm \ref{algo:1}. The proof is by induction. \\

\noindent We begin with the probability distribution of $k_1$. We have
    \begin{eqnarray*}
    Pr(k_1=1) & = & Pr(I_1=1) = \pi_1(0) = a_1^1.
    \end{eqnarray*}
Also, for $k \in \{1,\ldots,N-n+1\}$, we have
    \begin{eqnarray*}
    Pr(k_1=k) & = & Pr(I_1=\ldots=I_{k-1}=0,I_k=1) \\
              & = & \left\{\prod_{l=1}^{k-1} \left(1-n' \frac{\pi_l(0)}{n'-\pi_{l-1}^{+}(0)} \right) \right\}
              \left\{n' \frac{\pi_k(0)}{n'-\pi_{k-1}^{+}(0)}\right\} \\
              & = & \frac{\prod_{l=1}^{k-1} (n'-\pi_{l-1}^{+}(0)-n'\pi_l(0))}
                         {n'\left\{\prod_{l=2}^{k-1} (n'-\pi_{l-1}^{+}(0)) \right\} \{n'-\pi_{k-1}^{+}(0)\}} \{n' \pi_k(0)\} \\
              & = & \frac{\prod_{l=1}^{k-1} (n'-\pi_{l-1}^{+}(0)-n'\pi_l(0))}
                         {\prod_{l=1}^{k-1} (n'-\pi_{l}^{+}(0))} \{\pi_k(0)\} \\
              & = & \left[\prod_{l=1}^{k-1} \left\{1-(n'-1)\frac{\pi_l(0)}{n'-\pi_l^{+}(0)} \right\}\right] \times \pi_k(0) = a_k^1.
    \end{eqnarray*}
From equations (\ref{sec1:eq5}) and (\ref{sec1:eq6}), $i_1$ and $k_1$ have the same distribution. \\

\noindent Now, suppose that units $k_1,\ldots,k_{j-1}$ have been selected. We have
    \begin{eqnarray*}
    Pr(k_j=k_{j-1}+1|k_1,\ldots,k_{j-1})
    & = & Pr(I_{k_{j-1}+1}=1|k_1,\ldots,k_{j-1}) \\
    & = & Pr(I_{k_{j-1}+1}=1|n'_{k_{j-1}}=j-1) \\
    & = & (n'-j+1) \frac{\pi_{k_{j-1}+1}(0)}{n'-\pi_{k_{j-1}}^{+}(0)} \\
    & = & \left\{\frac{n'}{n'-\pi_{k_{j-1}}^{+}(0)} \right\} a_{k_{j-1}+1}^{j}.
    \end{eqnarray*}
Also, for $k \in \{k_{j-1}+2,\ldots,N-n+j\}$, we have
    \begin{eqnarray*}
    Pr(k_j=k|k_1,\ldots,k_{j-1})
    & = & Pr(I_{k_{j-1}+1}=\ldots=I_{k-1}=0,I_k=1|n'_{k_{j-1}}=j-1) \\
    & = & \left\{\prod_{l=k_{j-1}+1}^{k-1} \left(1-(n'-j+1) \frac{\pi_l(0)}{n'-\pi_{l-1}^{+}(0)} \right) \right\}
    \left\{(n'-j+1) \frac{\pi_k(0)}{n'-\pi_{k-1}^{+}(0)}\right\} \\
    & = & \frac{\prod_{l=k_{j-1}+1}^{k-1} \{n'-\pi_{l-1}^{+}(0)-(n'-j+1)\pi_l(0)\}}
                         {\{n'-\pi_{k_{j-1}}^{+}(0)\}\left\{\prod_{l=k_{j-1}+2}^{k-1} (n'-\pi_{l-1}^{+}(0)) \right\} \{n'-\pi_{k-1}^{+}(0)\}} \{n'-j+1\} \pi_k(0) \\
    & = & \frac{\prod_{l=k_{j-1}+1}^{k-1} \{n'-\pi_{l-1}^{+}(0)-(n'-j+1)\pi_l(0)\}}
               {\prod_{l=k_{j-1}+1}^{k-1} (n'-\pi_{l}^{+}(0))}  \times \frac{n'-j+1}{n'-\pi_{k_{j-1}}^{+}(0)} \pi_k(0) \\
    & = & \left[\prod_{l=k_{j-1}+1}^{k-1} \left\{1-(n'-j)\frac{\pi_l(0)}{n'-\pi_l^{+}(0)} \right\}\right] \times \frac{n'-j+1}{n'-\pi_{k_{j-1}}^{+}(0)} \pi_k(0) \\
    & = & \left\{\frac{n'}{n'-\pi_{k_{j-1}}^{+}(0)} \right\} a_{k}^{j}.
    \end{eqnarray*}
From equations (\ref{sec1:eq5}) and (\ref{sec1:eq6}), $i_j$ and $k_j$ have the same conditional distribution. This completes the proof.

\section{Proof of proposition \ref{prop:incons}} \label{app:incons}

\noindent We note $\Delta=\Delta_1+\Delta_2$, where
    \begin{eqnarray}
    \Delta_1 & = & \left\{\frac{n}{\pi_{N-n}^+ +n\pi_{N-n+1}}-1\right\} \sum_{k=1}^{N-n} y_k, \\
    \Delta_2 & = & \sum_{k=N-n+1}^{N} \frac{y_k}{\pi_k} \left\{\frac{n\pi_{N-n+1}}{\pi_{N-n}^+ +n\pi_{N-n+1}}-\pi_k \right\}.
    \end{eqnarray}
By using equation (\ref{sec4:eq5}) and the inequality
    \begin{eqnarray*}
    \frac{n}{\pi_{N-n}^+ +n\pi_{N-n+1}}-1 & = & \frac{\sum_{k=N-n+1}^N \pi_k - n\pi_{N-n+1}}{\pi_{N-n}^+ +n\pi_{N-n+1}} \\
    & \geq & \frac{\sum_{k=N-n+2}^N (\pi_k - \pi_{N-n+1})}{n} \\
    & = & \frac{1}{n} \sum_{i=1}^{n-1} (n-i)(\pi_{N-n+i+1}-\pi_{N-n+i}),
    \end{eqnarray*}
we first obtain
    \begin{eqnarray}
    |\Delta_1| & \geq & \lambda_2 \frac{n}{N} \left|\sum_{k=1}^{N-n} y_k\right|.
    \end{eqnarray}
Also, from the assumption (SD1) and from the inequality
    \begin{eqnarray*}
    \left|\frac{n\pi_{N-n+1}}{\pi_{N-n}^+ +n\pi_{N-n+1}} \right| & \leq & \frac{n}{N} \frac{\pi_N}{\pi_1},
    \end{eqnarray*}
we obtain
    \begin{eqnarray}
    |\Delta_2| & \leq & \frac{\Lambda_1}{\lambda_1} \left(1+\frac{1}{\lambda_1}\right) \sum_{k=N-n+1}^N |y_k|
    \end{eqnarray}
This leads to
    \begin{eqnarray}
    |\Delta| \geq |\Delta_1|-|\Delta_2| \geq n \left\{\lambda_2 (1-f) c_2 - \frac{\Lambda_1}{\lambda_1} \left(1+\frac{1}{\lambda_1}\right) C_2 \right\},
    \end{eqnarray}
and the result follows from equation (\ref{prop:incons:eq2}) and from the inequality     \begin{eqnarray*}
    \delta_n & = & (1-\pi_N) \frac{\pi_{N-n}^+ +n\pi_{N-n+1}}{\pi_{N-n}^+} \geq (1-\pi_N).
    \end{eqnarray*}

\section{Proof of proposition \ref{prop:cons}} \label{app:cons}

\noindent Making use of Theorem 3 in \cite{vij:68}, we have
    \begin{eqnarray} \label{app:cons:eq1}
    V\{\hat{t}_{y\pi}|\pi(0)\} & \leq & \sum_{k \in U'} \pi_k(0) \left\{\frac{y_k}{\pi_k}-\frac{1}{n'} \sum_{l \in U'} \frac{y_k}{\pi_k} \right\}^2 \leq \sum_{k \in U} \pi_k(0) \left\{\frac{y_k}{\pi_k}\right\}^2 \nonumber \\
    \Rightarrow EV\{\hat{t}_{y\pi}|\pi(0)\} & \leq &  \sum_{k \in U} \frac{y_k^2}{\pi_k},
    \end{eqnarray}
and from assumptions (VA1) and (SD1), $EV\{\hat{t}_{y\pi}|\pi(0)\}=O(N^2 n^{-1})$. \\

\noindent We also have
    \begin{eqnarray} \label{app:cons:eq2}
    VE\{\hat{t}_{y\pi}|\pi(0)\} = V\{\xi(0)\} & = & \sum_{i=1}^{n-1} \delta_i E\{\xi(0)^2|n'=i\}+\delta_n E\{\xi(0)^2|n'=n\}.
    \end{eqnarray}
For $i<n$, we obtain from the assumptions that $\delta_i=o(N^{-1})$ and $E\{\chi(0)^2|n'=i\}=O(N^4 n^{-2})$, so that the first term in the rhs of (\ref{app:cons:eq2}) is $o(N^3 n^{-1})=o(N^2)$. We can also write
    \begin{eqnarray} \label{app:cons:eq3}
    E\{\chi(0)^2|n'=n\} & = & \left[\left\{\frac{n}{\pi_{N-n}^+ +n\pi_{N-n+1}}-1\right\} \left\{\sum_{k=1}^{N-n} y_k + \pi_{N-n+1} \sum_{k=N-n+1} \frac{y_k}{\pi_k} \right\} \right. \nonumber \\
    & - & \left. \sum_{k=N-n+1}^N \frac{y_k}{\pi_k} (\pi_{k}-\pi_{N-n+1}) \right]^2 \nonumber \\
    & \leq & 2 \left\{\frac{n}{\pi_{N-n}^+ +n\pi_{N-n+1}}-1\right\}^2 \left\{\sum_{k=1}^{N-n} y_k + \pi_{N-n+1} \sum_{k=N-n+1} \frac{y_k}{\pi_k} \right\}^2 \nonumber \\
    & + & 2 \left\{\sum_{k=N-n+1}^N \frac{y_k}{\pi_k} (\pi_{k}-\pi_{N-n+1}) \right\}^2.
    \end{eqnarray}
From the identity
    \begin{eqnarray*}
    \frac{n}{\pi_{N-n}^+ +n\pi_{N-n+1}}-1 & = & \frac{\sum_{k=N-n+2}^N (\pi_k-\pi_{N-n+1})}{\pi_{N-n}^+ +n\pi_{N-n+1}} = \frac{\sum_{i=1}^{n-1} (n-i)(\pi_{N-n+i+1}-\pi_{N-n+i})}{\pi_{N-n}^+ +n\pi_{N-n+1}}
    \end{eqnarray*}
and from the assumptions, the first term in the rhs of (\ref{app:cons:eq3}) is $o(n^2)$, while the second term in the rhs of (\ref{app:cons:eq3}) is $o(N^2)$. From (\ref{app:cons:eq2}), we obtain that $VE\{\hat{t}_{y\pi}|\pi(0)\}=o(N^2)$. This completes the proof.

\section{Proof of Proposition \ref{prop3}} \label{appB}

\subsection*{Preliminary result}

\begin{Lemma} \label{lem1}
Suppose that Assumptions (SD1) and (SD2) hold. Then some constants $C_3$ and $C_4$ exist such that
    \begin{eqnarray*}
    E\left(\frac{1}{n'}\right) & \leq & C_3 h(n,N) \ln(n) + \frac{C_4}{N}.
    \end{eqnarray*}
\end{Lemma}

\subsubsection*{Proof}

%\begin{propproof}
\noindent We have
    \begin{eqnarray*}
    E\left(\frac{1}{n'}\right) & = & \sum_{i=1}^n \frac{\delta_i}{i} \\
                               & = & \sum_{i=1}^n \left(\pi_{N-n+i+1}-\pi_{N-n+i}\right) \frac{\pi_{N-n}^+ + i \pi_{N-n+1}}{\pi_{N-n}^+} \frac{1}{i} \\
                               & = & \sum_{i=1}^n \frac{\pi_{N-n+i+1}-\pi_{N-n+i}}{i} + \frac{\pi_{N-n+1}(1-\pi_{N-n+1})}{\pi_{N-n}^+},
    \end{eqnarray*}
which gives the result.
%\end{propproof}

\subsection*{Proof of Proposition \ref{prop3}}

\noindent First note that from equation (\ref{sec5:eq3}), we have
    \begin{eqnarray*}
    V\left\{\hat{t}_{y\pi}(0)\right\} & = & EV\left\{\hat{t}_{y\pi}(0)|\pi(0)\right\}.
    \end{eqnarray*}
By using Theorem 3 in \cite{vij:68}, we have
    \begin{eqnarray*}
    V\left\{\hat{t}_{y\pi}(0)|\pi(0)\right\} & = & -\frac{1}{2} \sum_{k \neq l \in U'} \left\{\pi_{kl}(0)-\pi_{k}(0)\pi_{l}(0) \right\} \left\{\frac{y_k}{\pi_k(0)}-\frac{y_l}{\pi_l(0)}\right\}^2 \\
    & \leq & \frac{1}{2n'} \sum_{k \neq l \in U'} \pi_{k}(0)\pi_{l}(0) \left\{\frac{y_k}{\pi_k(0)}-\frac{y_l}{\pi_l(0)}\right\}^2 \\
    & = & \sum_{k \in U'} \pi_k(0) \left\{\frac{y_k}{\pi_k(0)} - \frac{1}{n'} \sum_{l \in U'} y_k \right\}^2 \\
    & \leq & \sum_{k \in U'} \frac{y_k^2}{\pi_k(0)}.
    \end{eqnarray*}
By using the inequality
    \begin{eqnarray}
    \pi_{N-n}^+ + n' \pi_{N-n+1} & \leq & \pi_{N-n}^+ + n \pi_{N-n+1} \leq \pi_N^+ = n,
    \end{eqnarray}
we obtain under assumption (H1) that for any $k \in U'$:
    \begin{eqnarray} \label{ineq:pik:0}
    \pi_{k}(0) & \geq & f_0 \frac{n'}{N},
    \end{eqnarray}
and the result follows from Assumption (H2) and Lemma \ref{lem1}.

%\bibliographystyle{apalike}
%\bibliography{biblio_gen}
%%\bibliography{C:/Users/Dell-000094/Dropbox/RedactionArticles/_Biblio/biblio_gen}

\begin{thebibliography}{}

\end{thebibliography}


\begin{thebibliography}{}

\bibitem[Chaudhuri and Vos, 1988]{cha:vos:88}
Chaudhuri, A. and Vos, J. (1988).
\newblock Unified theory and strategies of survey sampling.
\newblock Technical report.

\bibitem[Chauvet, 2020]{cha:20}
Chauvet, G. (2020).
\newblock A note on chromy's sampling procedure.
\newblock {\em Journal of Survey Statistics and Methodology}.

\bibitem[Chauvet and Vall{\'e}e, 2020]{cha:val:20}
Chauvet, G. and Vall{\'e}e, A.-A. (2020).
\newblock Inference for two-stage sampling designs.
\newblock {\em Journal of the Royal Statistical Society: Series B (Statistical
  Methodology)}, 82(3):797--815.

\bibitem[Chromy, 1979]{chr:79}
Chromy, J.~R. (1979).
\newblock Sequential sample selection methods.
\newblock In {\em Proceedings of the Survey Research Methods Section of the
  American Statistical Association}, pages 401--406.

\bibitem[Deville and Till{\'e}, 1998]{dev:til:98}
Deville, J.-C. and Till{\'e}, Y. (1998).
\newblock Unequal probability sampling without replacement through a splitting
  method.
\newblock {\em Biometrika}, 85(1):89--101.

\bibitem[Fan et~al., 1962]{fan:mul:rez:62}
Fan, C., Muller, M.~E., and Rezucha, I. (1962).
\newblock Development of sampling plans by using sequential (item by item)
  selection techniques and digital computers.
\newblock {\em Journal of the American Statistical Association},
  57(298):387--402.

\bibitem[Hanurav, 1967]{han:67}
Hanurav, T. (1967).
\newblock Optimum utilization of auxiliary information: $\pi$ps sampling of two
  units from a stratum.
\newblock {\em Journal of the Royal Statistical Society: Series B
  (Methodological)}, 29(2):374--391.

\bibitem[Jang et~al., 2010]{jan:pro:she:10}
Jang, L., Provost, M., and Sherk, A. (2010).
\newblock Challenges in the design of the canadian community health survey on
  healthy aging.
\newblock In {\em Proceedings of the Joint Statistical Meetings, American
  Statistical Association}, pages 2452--2466.

\bibitem[Kulathinal et~al., 2007]{kul:kar:saa:kuu:07}
Kulathinal, S., Karvanen, J., Saarela, O., and Kuulasmaa, K. (2007).
\newblock Case-cohort design in practice--experiences from the morgam project.
\newblock {\em Epidemiologic Perspectives \& Innovations}, 4(1):1--17.

\bibitem[Langlet et~al., 2003]{lan:fau:les:03}
Langlet, {\'E}.~R., Faucher, D., and Lesage, {\'E}. (2003).
\newblock An application of the bootstrap variance estimation method to the
  canadian participation and activity limitation survey.
\newblock In {\em Proceedings of the Joint Statistical Meetings, American
  Statistical Association}, pages 2299--2306.

\bibitem[Myrskyl{\"a}, 2007]{myr:07}
Myrskyl{\"a}, M. (2007).
\newblock {\em Generalised regression estimation for domain class frequencies}.
\newblock PhD thesis, University of Helsinki.

\bibitem[Sampford, 1967]{sam:67}
Sampford, M. (1967).
\newblock On sampling without replacement with unequal probabilities of
  selection.
\newblock {\em Biometrika}, 54(3-4):499--513.

\bibitem[Sunter, 1977]{sun:77}
Sunter, A. (1977).
\newblock List sequential sampling with equal or unequal probabilities without
  replacement.
\newblock {\em Journal of the Royal Statistical Society: Series C (Applied
  Statistics)}, 26(3):261--268.

\bibitem[Sunter, 1986]{sun:86}
Sunter, A. (1986).
\newblock Solutions to the problem of unequal probability sampling without
  replacement.
\newblock {\em International Statistical Review/Revue Internationale de
  Statistique}, pages 33--50.

\bibitem[Till{\'e}, 2011]{til:06}
Till{\'e}, Y. (2011).
\newblock {\em Sampling algorithms}.
\newblock Springer.

\bibitem[Vijayan, 1968]{vij:68}
Vijayan, K. (1968).
\newblock An exact $\pi$-ps sampling scheme—generalization of a method of
  hanurav.
\newblock {\em Journal of the Royal Statistical Society: Series B
  (Methodological)}, 30(3):556--566.

\bibitem[Xiong and Higgins, 2020]{xio:hig:20}
Xiong, Y. and Higgins, M.~J. (2020).
\newblock The benefits of probability-proportional-to-size sampling in
  cluster-randomized experiments.
\newblock {\em arXiv preprint arXiv:2002.08009}.

\bibitem[Zhao, 2011]{zha:11}
Zhao, Y. (2011).
\newblock Estimating the size of an injecting drug user population.
\newblock {\em World Journal of AIDS}, 1(03):88.

\end{thebibliography}

\end{document}